\documentclass[prb,twocolumn,showpacs,preprintnumbers,amsmath,amssymb]{revtex4}

\usepackage{graphicx}
\usepackage{dcolumn}
\usepackage{bm}

\begin{document}

\title{Dependence of quantum-Hall conductance on the edge-state
equilibration position in a bipolar graphene sheet}

\author{Dong-Keun Ki$^1$, Seung-Geol Nam$^1$, Hu-Jong
Lee$^{1,2}$,} \altaffiliation{Corresponding author:
hjlee@postech.ac.kr} \author{Barbaros \"{O}zyilmaz$^3$}

\affiliation{$^1$ Department of Physics, Pohang University of
Science and Technology, Pohang 790-784, Republic of Korea \\
$^2$National Center for Nanomaterials Technology, Pohang 790-784,
Republic of Korea \\ $^3$Department of Physics, National
University of Singapore, Singapore 117542}

\date{\today}

\begin{abstract}
By using four-terminal configurations, we investigated the
dependence of longitudinal and diagonal resistances of a graphene
p-n interface on the quantum-Hall edge-state equilibration
position. The resistance of a p-n device in our four-terminal
scheme is asymmetric with respect to the zero point where the
filling factor ($\nu$) of the entire graphene vanishes. This
resistance asymmetry is caused by the chiral-direction-dependent
change of the equilibration position and leads to a deeper insight
into the equilibration process of the quantum-Hall edge states in
a bipolar graphene system.
\end{abstract}

\pacs{73.43.Fj, 71.70.Di, 73.61.Wp, 73.23.-b}

\maketitle

\begin{figure}[t]
\includegraphics[width=8.5cm]{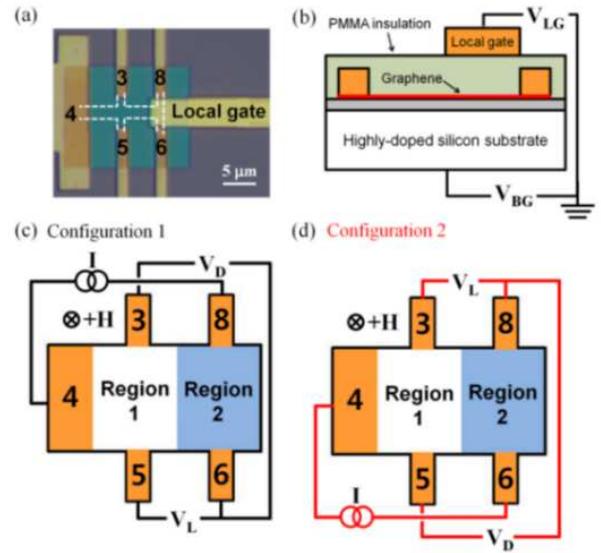}
\caption{(Color online) (a) Optical image of the sample. A PMMA
insulation layer covers the graphene sheet, the boundary of which
is defined by white broken lines. The white scale bar represents 5
$\mu$m. (b) Schematic gate configuration. (c) Measurement
configuration 1. (d) Measurement configuration 2.}
\end{figure}

\begin{figure}[t]
\includegraphics[width=8.5cm]{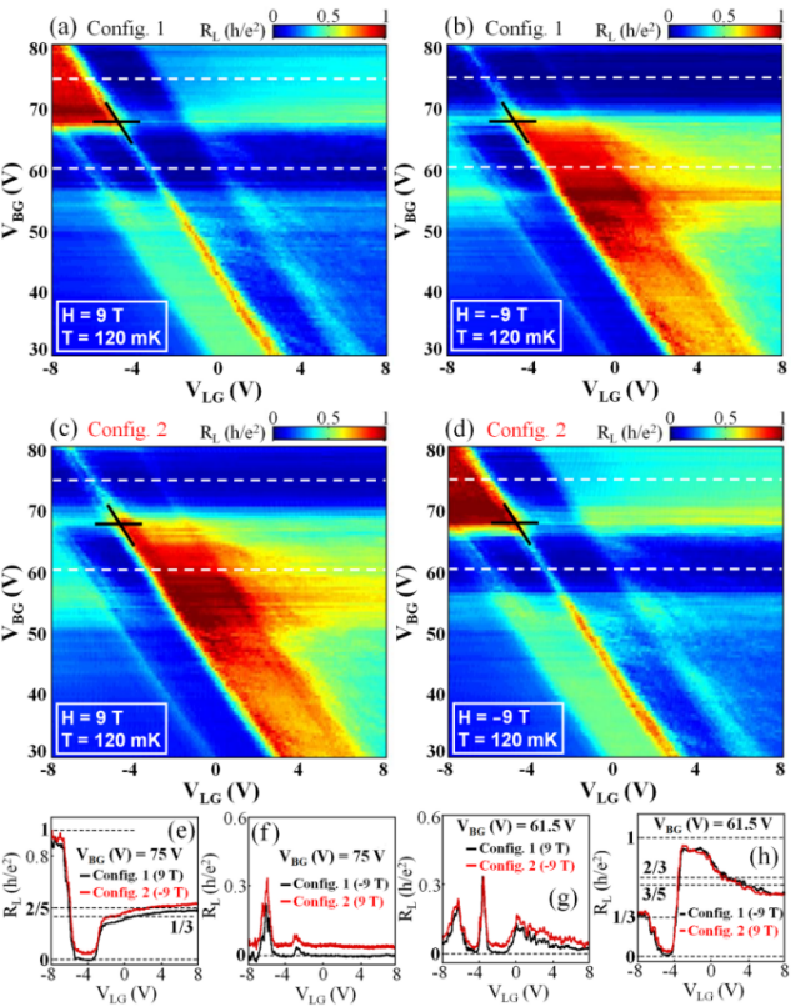}
\caption{(Color online) (a,b) 2D contrast maps of
$R_L$($V_{LG}$,$V_{BG}$) for the configuration 1, and $H$=9 T and
-9 T, respectively. (c,d) 2D contrast maps of
$R_L$($V_{LG}$,$V_{BG}$) for the configuration 2, and $H$=9 T and
-9 T, respectively. White broken lines in (a-d) indicate the
values of $V_{BG}$ where the slice data in (e-h) are extracted.
(e,f) Slice plots of 2D contrast maps at $V_{BG}$=75 V. (g,h)
Slice plots of 2D contrast maps at $V_{BG}$=61.5 V. Black and gray
lines in (e-h) represent the slice data for the configurations 1
and 2, respectively.}
\end{figure}

\begin{figure}[t]
\includegraphics[width=8.5cm]{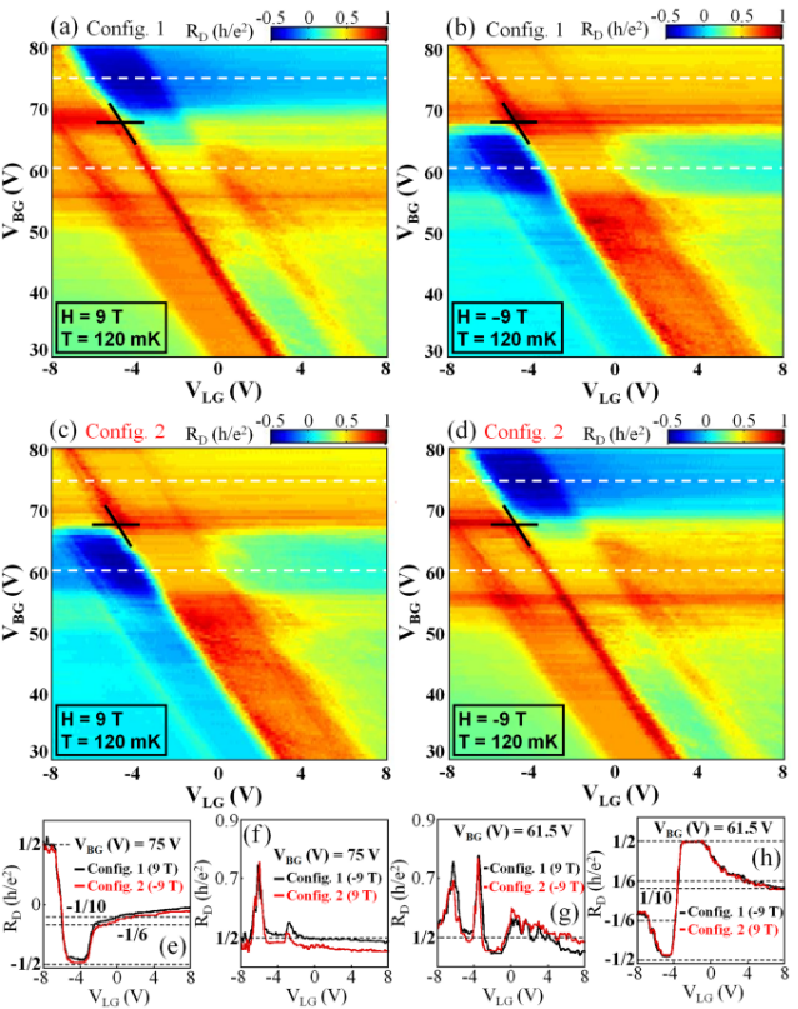}
\caption{(Color online) (a,b) 2D contrast maps of
$R_D$($V_{LG}$,$V_{BG}$) for the configuration 1, and $H$=9 T and
-9 T, respectively. (c,d) 2D contrast maps of
$R_D$($V_{LG}$,$V_{BG}$) for the configuration 2, and $H$=9 T and
-9 T, respectively. White broken lines in (a-d) indicate the
values of $V_{BG}$ where slice data in (e-h) are extracted. (e,f)
Slice plots of 2D contrast maps at $V_{BG}$=75 V. (g,h) Slice
plots of 2D contrast maps at $V_{BG}$=61.5 V. Black and gray lines
in (e-h) represent the slice data for the configurations 1 and 2,
respectively.}
\end{figure}

\begin{figure}[t]
\includegraphics[width=8.5cm]{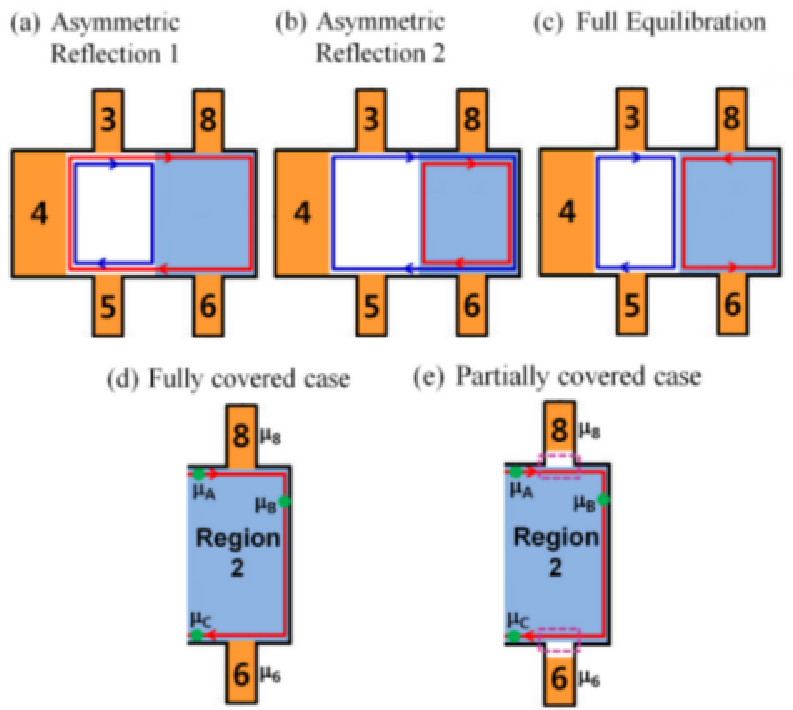}
\caption{(Color online) Schematic configurations of the edge-state
circulation for (a) the asymmetric reflection 1, (b) the the
asymmetric reflection 2, and (c) the full equilibration. The
configuration with the local gate (d) fully covering and (e)
partially covering the area between leads 6 and 8.}
\end{figure}

Half-integer quantum-Hall (QH) effect in graphene manifests the
massless-Dirac-fermionic nature of carriers in the
material.~\cite{Novoselov05,Zhang05} The corresponding QH edge
state in graphene has been explicitly confirmed to be
chiral~\cite{Ki09-1} as in an ordinary two-dimensional (2D)
electron gas.~\cite{Halperin82,Buttiker88,Beenakker91} Since
carriers in graphene possess bipolar characteristics, the filling
factor ($\nu$) and the chiral direction of the edge states can be
modulated by gating a graphene sheet
electrostatically.~\cite{Novoselov05,Zhang05} Although the spatial
deflection of an edge state is hard to be
realized,~\cite{Katsnelson06} a partial reflection of the edge
state can be accomplished in locally gated graphene devices by
means of the edge-state equilibration at p-n
interfaces.~\cite{Williams07,Abanin07,Ozyilmaz07,Ki09}
Nonetheless, previous conductance measurements, either in
two-terminal~\cite{Williams07,Abanin07,Ozyilmaz07} or in
four-terminal configurations,~\cite{Ki09} did not explicitly
reveal the dependence of the conductance on the equilibration
position (EP), where the edge states with different filling
factors are equilibrated.

Recently, by measuring four-terminal magnetoresistance of a
spatially chemical-doped graphene p-n device, the longitudinal
resistances are shown to be either quantized at finite values or
vanish depending on the gate voltage, the measurement
configuration, and the magnetic field ($H$)
direction.~\cite{Lohmann09} Analysis of the data confirmed that
the four-terminal magnetoresistance depends on the location of the
EP. From this result, one may suppose that the location of the EP
can be experimentally determined simply by measuring four-terminal
longitudinal magnetoresistance in any hybrid systems. In the
study, however, the $\nu$ of the chemically doped region was not
flexibly controllable, as distinct from the locally top-gated
devices, where the $\nu$ of the top-gated region can be controlled
separately from the region of the global gating only. In this
study, we took the longitudinal ($R_L$) and the diagonal ($R_D$)
resistances of a graphene p-n interface established by a local
top-gate in four-terminal configurations. The measurement scheme
enabled us to access in detail the development of the
equilibration condition of the QH edge conducting states in
graphene. We locate the EP's corresponding to different
measurement configurations and explicitly show that both
four-terminal $R_L$ and $R_D$ are sensitive to the location of the
EP.

A graphene p-n device was fabricated by electron-beam patterning
and depositing a Cr (5 nm)/Au (25 nm) bilayer on a mechanically
exfoliated mono-layer graphene sheet.~\cite{Novoselov04} It was
then followed by the polymethyl methacrylate (PMMA; 950 K, 2{\%}
in anisole) insulation~\cite{Huard07} and the deposition of a
local-gate 
, which covered one side of the sample~\cite{Ki09} [Figs. 1(a,b)].
The device was cooled down to 120 mK in a dilution fridge and
$R_D$ (=$V_D$/$I$) and $R_L$ (=$V_L$/$I$) were obtained in two
different measurement configurations [Figs. 1(c,d)] by using the
conventional lock-in technique.~\cite{Ki09} The bipolar character
of the device was verified at zero magnetic field $H$ by measuring
$R_L$ as a function of local-gate and back-gate voltages ($V_{LG}$
and $V_{BG}$, respectively), which exhibits a cross-like structure
in the parameter space set up by the local variation of the
density and the type of conduction carriers~\cite{Ki09} (not
shown).

The 2D contrast maps in Fig. 2 display $R_L$($V_{LG}$,$V_{BG}$)
measured in two different configurations at $H$=$\pm$9 T. Figs.
2(a-d) have several skewed blocks with varied contrast, i.e., with
differently quantized resistances. Each figure has an asymmetry
with respect to the zero point (a black cross), where the value of
$\nu$ of the QH edge states in the regions 1 ($\nu_1$) and 2
($\nu_2$) vanishes. The region 1 (2) represents the area outside
(underneath) the local gate [Figs. 1(c,d)]. Figure 2(a), which
corresponds to the configuration 1 and $H$=+9 T, is analogous to
Fig. 2(d) for the configuration 2 and $H$=$-$9 T. Similarly, Fig.
2(b) nearly duplicates Fig. 2(c). On the other hand, resistances
for opposite configurations at a given $H$ [Figs. 2(a,b) or Figs.
2(c,d)] have 180$^\circ$-rotational symmetry with respect to the
zero point. This is in contrast to earlier two-terminal studies,
where conductances were always
symmetric,~\cite{Williams07,Ozyilmaz07} as well as to the
four-terminal studies on a p-n-p device, where only $R_D$
exhibited an asymmetry.~\cite{Ki09}

Details of resistance variations can be analyzed by examining
slices of Figs. 2(a-d) at two different values of $V_{BG}$, as
shown in Figs. 2(e-h). Each figure contains two such slices black
and gray lines for a concurrent change of the measurement
configuration and the $H$ direction. As expected, black and gray
lines in each slice are almost identical. $R_L$'s in Figs. 2(f,g)
oscillate with zero resistance plateaus, which are the
Shubnikov-de Hass oscillations in the region 2. However, $R_L$'s
in Figs. 2(e,h) show a few plateaus, some with fractional
resistances. This is an indication of the finite edge-state
reflection at the p-n interface.~\cite{Ki09} One notes that the
difference between Fig. 2(e) [Fig. 2(h)] and Fig. 2(f) [Fig.
2(g)], which are in an identical configuration, is in the $H$
direction only. It will be shown below that this seeming strange
feature arises from the sensitivity of a four-terminal resistance
on the location of the EP in our graphene p-n device.

We also measured $R_D$($V_{LG}$, $V_{BG}$) as plotted in Figure 3.
Similar to the behavior of Figs. 2(a-d), Fig. 3(a) [Fig. 3(b)]
resembles Fig. 3(d) [Fig. 3(c)] but with an inversion symmetry
with Fig. 3(c) [Fig. 3(d)]. Each slice plot in Figs. 3(e-h) and
the corresponding one in Figs. 2(e-h) are qualitatively similar
despite some quantitative differences. Moreover, in Figs. 3(e-h),
one can see that $R_D$ is always the Hall resistance either for
the region 1 ($R_{H1}$) or for the region 2 ($R_{H2}$). It is
easily explained by the Kirchhoff's current-conservation relation,
which points out that $R_D$ is the summation of $R_{H1}$ or
$R_{H2}$ and the corresponding $R_L$. Consequently, symmetries
imposed upon $R_L$ should appear in $R_D$.

Transport properties of the QH edge state can be conveniently
analyzed by the Landauer-B\"{u}ttiker
formula.~\cite{Buttiker88,Beenakker91,Ki09} Based on the
complete-mode-mixing hypothesis,~\cite{Abanin07,Ki09} one can
distinguish three equilibration regimes: the asymmetric reflection
1 ($\nu_1$$\cdot$$\nu_2$$>$0 and $|\nu_1|$$\geq$$|\nu_2|$), the
asymmetric reflection 2 ($\nu_1$$\cdot$$\nu_2$$>$0 and
$|\nu_1|$$<$$|\nu_2|$), and the full equilibration
($\nu_1$$\cdot$$\nu_2$$<$0) regimes. Figs. 4(a-c) illustrate the
schematics of the edge state circulation in each regime for
$\nu_1$$<$0 and $+H$. The asymmetric reflection 1 [Fig. 4(a)] is
the case where the edge states inside the region 2 are not
reflected at the interface. The situation is opposite for the
asymmetric reflection 2 [Fig. 4(b)]. In full equilibration regime
[Fig. 4(c)], the edge states in the regions 1 and 2 merge at the
one side of the sample [for example, the upper side in Fig. 4(c)]
and split at the other side [the lower side in Fig. 4(c)].

By evaluating scattering matrices, we calculated $R_L$ and $R_D$
as a function of $\nu_1$ and $\nu_2$ for different measurement
conditions and equilibration
regimes.~\cite{Buttiker88,Beenakker91,Ki09} Detailed results are
summarized below for $\nu_1$$<$0 and $+H$.
\\

(a) Asymmetric reflection 1 ($\nu_1$$\cdot$$\nu_2$$>$0,
$|\nu_1|$$\geq$$|\nu_2|$)
\begin{eqnarray}
R_D &=& \begin{cases} \frac{h}{e^2}\frac{1}{|\nu_2|} &
\text{(Configuration 1)} \\
-\frac{h}{e^2}\frac{1}{|\nu_1|} & \text{(Configuration 2)}
\end{cases}, \\
R_L &=& \begin{cases}
\frac{h}{e^2}\frac{|\nu_1|-|\nu_2|}{|\nu_1||\nu_2|} &
\text{(Configuration 1)} \nonumber \\
0 & \text{(Configuration 2)}
\end{cases}
\end{eqnarray}

(b) Asymmetric reflection 2 ($\nu_1$$\cdot$$\nu_2$$>$0,
$|\nu_1|$$<$$|\nu_2|$)
\begin{eqnarray}
R_D &=& \begin{cases} \frac{h}{e^2}\frac{1}{|\nu_1|} &
\text{(Configuration 1)} \\
-\frac{h}{e^2}\frac{1}{|\nu_2|} & \text{(Configuration 2)}
\end{cases}, \\
R_L &=& \begin{cases} 0 &
\text{(Configuration 1)} \nonumber \\
\frac{h}{e^2}\frac{|\nu_2|-|\nu_1|}{|\nu_1||\nu_2|} &
\text{(Configuration 2)}
\end{cases}
\end{eqnarray}

(c) Full equilibration ($\nu_1$$\cdot$$\nu_2$$<$0)
\begin{eqnarray}
R_D &=& \begin{cases} \frac{h}{e^2}\frac{1}{|\nu_1|} &
\text{(Configuration 1)} \\
\frac{h}{e^2}\frac{1}{|\nu_2|} & \text{(Configuration 2)}
\end{cases}, \\
R_L &=& \begin{cases} 0 &
\text{(Configuration 1)} \nonumber \\
\frac{h}{e^2}\frac{|\nu_1|+|\nu_2|}{|\nu_1||\nu_2|} &
\text{(Configuration 2)}
\end{cases}
\end{eqnarray}
\\

In Eq. (1), $R_L$ for the configuration 2 (=$V_{38}$/$I$) vanishes
but it is finite in Eqs. (2) and (3), which is in coincidence with
our experimental findings [compare Figs. 2(a) with 2(b) for
$V_{BG}$$<$$\sim$55 V]. Here, $V_{\alpha\beta}$ stands for the
voltage difference between the leads $\alpha$ and $\beta$. The
behavior of $R_L$ results because the EP in the asymmetric
reflection 1 regime is different from that in the other regimes.
In the former regime, the edge states are equilibrated at the
lower boundary of the sample between the leads 5 and 6 [Fig.
4(a)], so that $V_L$ (or equivalently $R_L$) for the configuration
1 (=$V_{56}$) is finite while the one for the configuration 2
(=$V_{38}$) vanishes. On the contrary, in the asymmetric
reflection 2 [Figs. 4(b)] and the full equilibration [Figs. 4(c)]
cases, the equilibration takes place at the upper boundary of the
sample between the leads 3 and 8, resulting in the finite $R_L$ in
the configuration 2 (=$V_{38}$/$I$). Since the EP depends on the
chiral direction of the edge-state circulation, $R_L$ in the
configuration 1 (2) for $\nu_1$$<$0 becomes identical to the one
in the configuration 2 (1) for $\nu_1$$>$0 for an identical $H$.
Equivalently, $R_L$ for $\nu_1$$<$0 in a negative $H$ becomes the
same as that for $\nu_1$$>$0 in a positive $H$. It agrees with
symmetries found in our experiment [Fig. 2]. Moreover, as shown as
broken lines in Figs. 2(e-h), calculations and experiments are in
a quantitative agreement with each other. It is important to point
out that, in a p-n-p device, the EP's at opposite sides of the
local-gate are located at different (upper or lower) boundaries of
the sample. Thus, four-terminal $R_L$ of a graphene p-n-p device
does not reveal the dependence on the location of EP, although the
EP itself sensitively depends on the chiral direction of the QH
edge state.~\cite{Ki09}

The calculated $R_D$ also coincides with the experimental value
[Fig. 3], which is always $R_{H1}$ or $R_{H2}$. This can be
explained by a simple Kirchhoff's law. For instance, $V_{36}$ is
the summation of $V_{38}$ and $V_{86}$(=$R_{H2}$$\times$$I$) or
equally $V_{35}$(=$R_{H1}$$\times$$I$) and $V_{56}$. As $V_{38}$
is zero in the asymmetric reflection 1 regime, $R_D$ for the
configuration 1 (=$V_{36}$/$I$) will be $R_{H2}$ [Eq. (1)]. For
the same reason, $V_{58}$/$I$ (=$R_D$ for the configuration 2)
should be -$R_{H1}$ [Eq. (1)]. In the other regimes, $V_{56}$
vanishes [Eqs. (2) and (3)], so that $R_D$ becomes $R_{H1}$ for
the configuration 1 and -$R_{H2}$ for the configuration 2.

Up to this point, we consider only the cases where all the regions
between the leads 6 and 8 are covered by local gates [Fig. 4(d)].
However, as seen in Fig. 4(e), the regions near the leads 6 and 8
may not be fully covered by the local gates, which results in the
formation of unintended p-n interfaces around the regions [dotted
squares in Fig. 4(e)]. This may change the conductance features we
discuss above. Apparently, the conductance change takes place in a
two-terminal configuration as the additional interface forms a
p-n-p junction between the source and the drain. But, in the
following discussion, we demonstrate that the results of
four-terminal measurements are not affected. This implies that the
difference in the chemical potential ($\mu$) between the point A
and the lead 6 ($\mu_A$$-$$\mu_6$) remains the same for both fully
and partially covered cases, when the leads 6 and 8 are a voltage
probe and a drain, respectively [here, we focus on the
configuration 1 in Fig. 1(c) but the same argument is valid for
the configuration 2]. As the chemical potential does not change
across the voltage probe without current flow, it is
straightforward to show that $\mu_6$ is always identical to
$\mu_B$ and $\mu_C$ for both cases. In addition, the Hall voltage
inside the region 2, $\mu_A$$-$$\mu_B$, does not vary irrespective
of the presence or the absence of p-n interfaces outside the
region 2 (the region covered by the local gate). Consequently, the
presence of additional bipolar interfaces does not affect the
value $\mu_A$$-$$\mu_6$, neither the corresponding four-terminal
resistances. One notes that, for the partially covered case,
$\mu_B$ can differ from $\mu_8$ in certain equilibration regimes
so that the partial coverage of the top gate affects the
two-terminal results. This demonstrates the high benefit of the
multi-terminal measurements in comparison with the two-terminal
ones for edge-state equilibration studies.

In this study, we independently controlled the local filling
factors of the top-gated and the rest of back-gated region of a
graphene sheet. It enabled us to access to diverse combination of
bipolar configurations in the sheet along with more systematic
examination of equilibration condition of QH edge conducting
states. We have shown that both the longitudinal ($R_L$ in Fig. 2)
and the diagonal ($R_D$ in Fig. 3) resistances, obtained in
four-terminal configurations, of this \emph{locally top-gated}
hybrid p-n junction device are asymmetric with respect to the zero
point. The behaviors of $R_L$ and $R_D$ are also in clear contrast
to the two-terminal studies on \emph{locally top-gated}
p-n~\cite{Williams07} and p-n-p~\cite{Ozyilmaz07} devices of
graphene as well as to similar four-terminal studies on
\emph{locally top-gated} p-n-p devices.~\cite{Ki09} The
resistances in two different measurement configurations [Figs.
1(c,d)] and $H$ directions confirms that the asymmetry is caused
by the dependence of the four-terminal resistance of a graphene
p-n interface on the location of the EP, which varies with the
chiral direction of the QH edge state. Since two-terminal
conductances do not depend on the chiral direction and the EP's in
a p-n-p device are always located at opposite boundaries of the
sample, the location of the EP could not be traced in the previous
experiments and the resistances turned out to be
symmetric.~\cite{Williams07,Ozyilmaz07,Ki09} This four-terminal
study on a graphene p-n interface provides a deeper insight into
how and where the QH edge-states are equilibrated in a locally
gated graphene system, which is essential in designing devices
such as quantum interferometers out of graphene.

\begin{acknowledgments}
This work was supported by National Research Foundation of Korea
through Acceleration Research Grant (No. R17-2008-007-01001-0) and
by the Ministry of Education, Science and Technology under the
grant No. 2009-0083380. BO acknowledges support from Singapore
National Research foundation under NRF Award No. NRF-RF2008-07 and
by NUS NanoCore.
\end{acknowledgments}

\end{document}